\documentclass[pdflatex,sn-mathphys-num]{sn-jnl}

\usepackage{graphicx}%
\usepackage{multirow}%
\usepackage{amsmath,amssymb,amsfonts}%
\usepackage{amsthm}%
\usepackage{mathrsfs}%
\usepackage[title]{appendix}%
\usepackage{xcolor}%
\usepackage{textcomp}%
\usepackage{manyfoot}%
\usepackage{booktabs}%
\usepackage{algorithm}%
\usepackage{algorithmicx}%
\usepackage{algpseudocode}%
\usepackage{listings}%
\usepackage{lmodern}
\usepackage{fix-cm}
\usepackage[T1]{fontenc} %
\usepackage{anyfontsize}

\begin{document}

\title[Shadows of black holes in dynamical Chern-Simons modified gravity]{Shadows of black holes in dynamical Chern-Simons modified gravity}

\author[]{\fnm{Benito} \sur{Rodríguez}}
\email{benito.rodriguez@fisica.uaz.edu.mx}

\author[]{\fnm{Javier} \sur{Chagoya}}
\email{javier.chagoya@fisica.uaz.edu.mx}

\author[]{\fnm{C.} \sur{Ortiz}}
\email{ortizgca@fisica.uaz.edu.mx}

\affil[]{\orgdiv{Unidad Académica de Física}, \orgname{Universidad Autónoma de Zacatecas}, \orgaddress{\street{Calzada Solidaridad esquina con Paseo a la Bufa S/N}, \postcode{98060}, \state{Zacatecas}, \country{México}}}

\abstract{We revisit and extend the study of null geodesics around a slowly rotating black hole in Chern-Simons modified gravity. We employ the Hamilton-Jacobi formalism to derive the equations for the shadow profile and determine its shape. We compare our results with numerical ray tracing, finding good agreement within the validity of our approximations for slow rotation and small Chern-Simons coupling. We forecast constraints on the
model parameters using the uncertainty in EHT data for the observed shadow of SgrA$^*$ as a reference.}

\maketitle

\tableofcontents
\clearpage

\section{Introduction}
In 1915, Albert Einstein formulated the General Theory of Relativity (GR), which relates gravity to the geometry of space-time. This theory has been extensively tested on the Solar System and other astrophysical scales~\cite{Will_2014} and is the basis of the Standard Cosmological Model. However; unless some exotic energy/matter content is considered, a few observations and theoretical questions elude GR's explanations, such as the late acceleration of the Universe~\cite{Perlmutter_1999,Riess_1998} and the anomalous rotation curves of galaxies~\cite{Sofue_2001}. 
Modifying GR offers alternative ways to address these unresolved issues, but any modification must reproduce well-tested predictions of GR. One such prediction, which has gained observational relevance in recent years, is the appearance of photons orbiting in unstable orbits around a black hole in an observer's sky, the black hole shadow~\cite{Stepanian_2021,Ayzenberg_2018,PhysRevD.106.064012,Antoniou_2023}.

A reason for the recent surge in black hole research is the reconstruction by the Event Horizon Telescope (EHT) of the first image of the shadow of the Supermassive Black Hole (SMBH) in the center of the $M87^*$ galaxy in 2019~\cite{Akiyama_2019}, followed by the image of the shadow of Sagittarius $A^{*}$, located in the center of the Milky Way~\cite{2022ApJ...930L..12E}, also reported by the EHT in 2022. This opens up new possibilities for constraining modified theories of gravity, since the size of the shadow can be affected both by the new coupling constants introduced by the alternative model of gravity and by the properties of the black hole solution, which usually is not the same as in GR.

In this work, we focus on shadows of black holes within modifications to GR that introduce a scalar field non-minimally coupled to squared curvature scalars, commonly known as quadratic gravity theories~\cite{Yunes_2011}. In particular, we work on the theory known as dynamical Chern-Simons gravity (dCS)~\cite{Jackiw_2003}. Several aspects of dCS have been explored in the literature. Cosmological studies have explored its connection with cosmic microwave background radiation~\cite{Lue_1999,Li_2007} and leptogenesis~\cite{Alexander_2006,Alexander_200}. Gravitational wave solutions have also been investigated in~\cite{Yunes_2010,Wagle_2022,Yagi_2012}.
Concerning black hole solutions in dCS~\cite{Wagle_2022,Grumiller_20008,PhysRevLett.28.1082,erdmenger2022universal,Kimura_2018}, 
spherically symmetric spacetimes in GR persist as solutions in dCS, whereas axially symmetric solutions do not. Furthermore, the dCS solution has the potential to result in violations of the Hawking-Penrose theorem~\cite{Alexander_2021}.
Yunes and Pretorius~\cite{Yunes_2009} reported the first solution that describes a rotating black hole in the slow rotation/small coupling limit. 
This solution introduces a modification to the Kerr metric parameterized by a scalar hair, resulting in a correction to the location of photon orbits. 
An analysis of these corrections was presented in~\cite{PhysRevD.81.124045,Vincent_2013, Meng_2023}. 
In this work, we revisit and expand on these results, presenting numerical shadows with an accretion disk and using observational data from M87$^*$ and SgrA$^*$ 
to put constraints on the parameters of the dCS spacetime.

 This paper is organized as follows. In Section~\ref{sec:cs} we present the basics of the dynamical Chern-Simons theory. In Section~\ref{sec:hj} we review the Hamilton-Jacobi formalism to investigate the shadow of the Kerr black hole in GR. 
In Section~\ref{sec:hjcs} we apply the Hamilton-Jacobi formalism to the slowly rotating black hole found in dynamical Chern-Simons gravity, and we also present images generated with the ray-tracing code \verb 'Gyoto'. We investigated the radius of the shadow and the distortion with respect to a Kerr black hole. These observable parameters allow us to compare the predictions of dCS with the observations reported by EHT.  We conclude in Section~\ref{sec:con} with a discussion of our results.

\section{Formulation of Chern-Simons modified gravity}\label{sec:cs}
Chern-Simons modified gravity (CS) represents a 4-dimensional deformation of GR defined by the action
\begin{equation}\label{actioncs}
S_\text{total}=S_\text{EH}+S_\text{CS}+S_{\vartheta}+S_\text{M}.
\end{equation}
The first term is the Einstein-Hilbert action,
\begin{equation}
S_\text{EH}= \kappa \int d^{4}x\sqrt{- g }R,
\end{equation}
where $\kappa=\left ( 16 \pi\right)^{-1}$ is the dimensionless gravitational coupling constant in units $c=G=1$, $R$ is the Ricci scalar, and $g$ is the determinant of the metric $g_{\mu\nu}$.

The CS term is given by
\begin{equation}
S_\text{CS} = \frac{\alpha}{4}\int d^{4}x \sqrt{-g}\vartheta {}^{\star}\!R R,
\end{equation}
where $\alpha$ is the CS coupling constant with units of length squared,  ${}^{\star}\!R$ is the dual Riemann tensor used to form the Pontryagin density ${}^{\star}\!RR$, and $\vartheta$ is a dimensionless scalar field, known as the Chern-Simons coupling field. In components,
\begin{eqnarray}\label{pontry}
{}^{\star}R R & = {}^{\star}R^{\mu}{}_{\nu}{}^{\rho \sigma} R^{\nu}{}_{\mu}{}_{\rho \sigma} = \frac{1}{2}\epsilon^{ \rho \sigma \delta\tau } R^{\mu}{}_{\nu}{}_{\delta\tau}R^{\nu}{}_{\mu}{}_{\rho \sigma}.
\end{eqnarray}
with $\epsilon^{ \rho \sigma \delta \tau }$ being the 4-dimensional Levi-Civita tensor,

The scalar field action is given by
\begin{equation}
S_{\vartheta}=-\frac{\beta}{2}\int d^{4}x\sqrt{-g}\left[g^{\mu\nu}\nabla_{\mu}\vartheta \nabla_{\nu}\vartheta +2V(\vartheta) \right],
\end{equation}
where $\nabla_{\mu}$ is the covariant derivative compatible with the metric, $\beta$ is a dimensionless coupling constant, and $V(\vartheta)$ is a potential for the scalar that we set to zero.

An additional, unspecified matter contribution is described by
\begin{equation}
S_\text{M}=\frac{1}{4}\int d^{4}x\sqrt{-g }\mathcal{L}_\text{M},
\end{equation}
where $\mathcal{L}_\text{M}$ is a Lagrangian density that does not depend on $\vartheta$. The modified CS equations include a variety of theories, each characterized by specific couplings $\alpha$ and $\beta$. Within this family, two distinct formulations stand out: the non-dynamical framework, where $\alpha$ can take any value while $\beta = 0$, and the dynamical framework, where both $\alpha$ and $\beta$ are arbitrary but nonzero. Unfortunately, since its inception, this theory has been explored with various notations for the coupling constants, here we adhere to the notation used in Ref.~\cite{Alexander_2009}.

If the scalar field $\vartheta$ is a constant, then the CS modified gravity is reduced identically to GR. This occurs because the Pontryagin term~\eqref{pontry} can be expressed as a divergence,
\begin{equation}
\nabla_{\mu}K^{\mu}=\frac{1}{2}{}^{\star}R R,
\end{equation}
 of the Chern-Simons topological current
\begin{equation}
    K^{\mu}=\epsilon^{ \mu \rho \sigma \tau }\Gamma^{\upsilon}_{\rho\psi}\left ( \partial_{\sigma}\Gamma^{\psi}_{\tau \upsilon}+\frac{2}{3}\Gamma^{\psi}_{\sigma\xi}\Gamma^{\xi}_{\tau \upsilon} \right ),
\end{equation}
where $\Gamma^{\psi}_{\sigma\xi}$ are the Christoffel symbols. 
Therefore, we focus on a non-constant $\vartheta$. 

The field equations are obtained by varying the action with respect to the metric and the CS coupling field. The former leads to
\begin{equation}
G_{\mu\nu}+\frac{\alpha}{\kappa}C_{\mu\nu}=\frac{1}{2\kappa}T_{\mu\nu},
\end{equation}
where $G_{\mu\nu}$ is the Einstein tensor and $C_{\mu\nu}$ is the trace-free C-tensor\footnote{Parenthesis around the indices denote symmetrization, e.g. $A^{(\mu\nu)}:=(A_{\mu\nu} + A_{\nu\mu})/2$.} 
\begin{equation}
    C^{\mu\nu} = \left( \nabla_{\sigma}\vartheta \right )\epsilon^{\sigma \delta \alpha ( \mu}
\nabla_{\alpha}R^{\nu)}{}_{\delta}+\left (\nabla_{\sigma}\nabla_{\delta}\vartheta \right ) {}^{\star}\!R^{\delta (\mu \nu )\sigma}.
\end{equation}
On the other hand, variation with respect to the CS coupling field leads to
\begin{equation}
    \beta\Box\vartheta=-\frac{\alpha}{4}\,{}^{\star}\!R R,
\end{equation}
where $\Box:=g^{\mu\nu}\nabla_{\mu}\nabla_{\nu}$ is the d’Alembert operator. The previous is the Klein-Gordon equation for a massless scalar field in the presence of a source term. 

\section{Hamilton-Jacobi Formalism in General Relativity}\label{sec:hj}

Much can be learned from the study of the geodesics of massive particles and photons around black holes, contributing to a deeper comprehension of the physics of space-time. For an analytical examination of these trajectories, the Hamilton-Jacobi formalism proves advantageous, and we provide a brief overview here. This will serve as a foundation for our subsequent exploration of geodesics in the dCS metric.

The space-time geometry in GR for a rotating black hole is described by the Kerr metric~\cite{PhysRevLett.11.237}. According to GR, the exterior region of stellar and supermassive black holes is characterized by this metric. In Boyer-Lindquist coordinates $(t, r, \theta, \phi)$, this metric takes the form:
\begin{align}\label{a0}
    ds^{2} = -\left(1 - \frac{2mr}{\rho^2} \right) dt^{2} 
    - \left(\frac{4mr\sin^{2}{\theta}}{\rho^2}\right) dtd\phi 
    + \left(\frac{\rho^2}{\Delta}\right) dr^2 
    + \rho^2 d\theta^2 \nonumber \\ 
    + \sin^{2}{\theta} \left(r^{2} + a^{2} + \frac{2mar\sin^{2}{\theta}}{\rho^2}\right) d\phi^{2}.
\end{align}
where
\begin{equation}
d\Omega=\left (r^{2}+a^{2}+\frac{2mar\sin^{2}{\theta}}{\rho^2}\right)d\phi ^{2},
\end{equation}
where $m$ is the mass in geometrized units
, and $\rho^2$ and $\Delta$ are defined as
\begin{equation}
\rho^2=r^{2}+a^{2}\cos^2{\theta},
\end{equation}
\begin{equation}
\Delta=r^{2}+a^{2}-2mr.
\end{equation}
The parameter $a$ measures the specific angular momentum~\cite{ryder_2009}. To prevent a naked singularity, observable from infinity and forbidden by the cosmic censorship conjecture~\cite{carroll1997lecture}, the value of $a$ is restricted to the interval $a\in [-m,m]$, or equivalently $\left| a \right|\leq m$.

\subsection{Hamilton-Jacobi equations in Kerr space-time}

The Hamiltonian for a particle in spacetime is expressed as
\begin{equation}
H=\frac{1}{2}g^{\mu\nu}p_{\mu}p_{\nu},
\end{equation}
where $p_\mu$ are the generalized momenta.
From the Hamilton-Jacobi formulation, this leads to 
\begin{equation}\label{HJ}
\frac{\partial S}{\partial \lambda}=-\frac{1}{2}g^{\mu\nu}\frac{\partial S}{\partial x^\mu}\frac{\partial S}{\partial x^\nu}.
\end{equation}
where $S$ is the Hamilton's principal function. 
In Kerr space-time, due to the symmetry of the problem, $H$ does not explicitly depend on $t$, $\phi$, and the affine parameter $\lambda$, resulting in three linear terms in $S$. Furthermore, it is assumed that the radial and angular dependencies are also separable. Then, the function $S$ is proposed as
\begin{equation}
S=\frac{\mu^2}{2}\lambda+Et+\Phi \phi+S_r(r)+S_\theta(\theta),\label{eq:hja}
\end{equation}
where
\begin{equation}
    \frac{\partial S}{\partial \lambda}=-H=\frac{\mu^2}{2},
\end{equation}
 and $\mu$ is the rest mass of the particle in geometrized units. The conserved quantities $E = -p_t$ and $\Phi =p_\varphi$ correspond to the specific energy and axial component of the angular momentum relative to an observer at infinity.
Substituting Eq.~\eqref{eq:hja} into the Hamilton-Jacobi Eq.~(\ref{HJ}) , we have
\begin{equation}
    g^{tt}E^2 +g^{\varphi\varphi}\Phi^{2}-2g^{t\varphi}E \Phi +g^{rr}\left ( \frac{dS_{r}}{dr} \right )^2+g^{\theta\theta}\left ( \frac{dS_{\theta}}{d\theta} \right )^2 =-\mu^{2}.
\end{equation}
Substituting the inverse terms of the Kerr metric and simplifying, we obtain
\begin{eqnarray}
    \frac{-P^{2}+p^{2}_{r}\Delta^{2}+\Delta r^{2}\mu^{2}}{\Delta} = & -\left[ a^{2}\mu^{2}\cos^{2}{\theta}+\left( aE\sin{\theta} - \frac{\Phi}{\sin\theta} \right)^2 +p^{2}_{\theta} \right],\label{eq:hjks}
\end{eqnarray}
where $P=E(r^2+a^2)-a\Phi $, $p_{r}={\partial S}/{\partial r}$, and $p_{\theta}={\partial S}/{\partial \theta}$.
The left hand side depends solely on the coordinate $r$ and is defined as $f_r(r)$, while the right hand side depends only on $\theta$ and is defined as $f_{\theta}(\theta)$. Therefore, the equality in Eq.~\eqref{eq:hjks} implies that both sides are equal to a constant, which we define as
\begin{equation}\label{K}
    f_{r}(r)=f_{\theta}(\theta)=-K.
\end{equation}
Sometimes, these equations are expressed in terms of the conserved quantity
\begin{equation}\label{Q}
    Q= K-(aE-\Phi)^{2},
\end{equation}
called Carter's constant. 
In terms of this quantity, it can be written as
\begin{equation}
    \Delta^{2}p^{2}_{r}=P^{2}-\Delta[Q+(aE-\Phi)^{2}+\mu^{2}r^{2}].
\end{equation}
Using $\dot{r}=p^{r}=g^{rr}p_{r}={\Delta} p_{r}/{\rho^{2}}$,
we obtain the equation of motion for $r$ as
\begin{equation}
    \rho^{2}\dot{r}=\pm\sqrt{P^{2}-\Delta[Q+(aE-\Phi)^{2}+\mu^{2}r^{2}]}=\pm\sqrt{R(r)}.\label{eq:kerrr}
\end{equation}
Similarly, the equation for $\theta$ is obtained from
$-K=f_\theta(\theta)$ and $\dot{\theta}=g^{\theta\theta}p_{\theta}={p_{\theta}}/{\rho^{2}}$, arriving to
\begin{equation}
    \rho^{2}\dot{\theta} = \sqrt{Q-\cos^{2}{\theta}\left ( a^{2}(\mu^{2}-E^{2})+\frac{\Phi^{2}}{\sin^{2}{\theta}} \right ) } =\sqrt{\Theta(\theta)}\label{eq:kerrt}.
\end{equation}
The orbits of the test particles in Kerr geometry are obtained from the solutions to Eqs.~\eqref{eq:kerrr}
and~\eqref{eq:kerrt}.
For photons, $\mu = 0$, the equations of motion for $r$ and $\theta$ become

\begin{align}
    \rho^4\dot{r}^{2} & =R(r)= P^{2}-\Delta\left ( Q+(aE-\Phi)^{2}+\mu^{2}r^{2}\right )\label{eq:RK},
\\
    \rho^4\dot{\theta}^{2} & = \Theta(\theta)=Q-\cos^{2}{\theta}\left ( a^{2}(\mu^{2}-E^{2})+\frac{\Phi^{2}}{\sin^{2}{\theta}} \right ).
\end{align}

Dividing the equation of motion for $r$ by $E^{2}$ and using the reduced Carter constants $\lambda={\Phi}/{E}, \eta={Q}/{E^{2}}$ 
we get
\begin{equation}\label{a2}
    \mathcal{R}(r)=\frac{R(r)}{E^{2}}=r^{4}+(a^{2}-\eta-\lambda^{2})r^{2}+2mr[\eta+(a-\lambda)^{2}]-\eta a^{2}.
\end{equation}
The conditions for spherical orbits~\cite{PhysRevD.5.814,Fathi,Teo_2021,Perlick_2017} are then given by 
\begin{align}
    \mathcal{R}(r) &= 0, \label{eq:r0} \\
    \frac{d\mathcal{R}(r)}{dr} &= 0.
\end{align}
The solution to this set of equations is provided by the roots of a quadratic equation. Each of these two roots leads to a different class of solutions, and only one of these classes corresponds to physically viable spherical orbits. In this class, it follows from Eq.~\eqref{eq:r0} that 
\begin{align}
    \lambda &= -\frac{r^{3} - 3mr^{2} + a^{2}r + ma^{2}}{a(r-m)}, \label{eqs:l} \\
    \eta &= -\frac{r^{3}(r^{3} - 6mr^{2} + 9m^{2}r - 4a^{2}m)}{a^{2}(r-m)^{2}}. \label{eqs:eta}
\end{align}
Even though $\lambda$ and $\eta$ are expressed in terms of the orbital radius, for a photon following a geodesic, these quantities remain constant. It is worth mentioning that these spherical orbits are unstable.

For Eqs.~\eqref{eqs:l} and~\eqref{eqs:eta}, the condition $\eta \geq 0$~\cite{Chandrasekhar:1985kt} restricts the range of the radial coordinate to $r_1 \leq r \leq r_2$, where $r_1$ and $r_2$ are the roots of $\eta$,
\begin{equation}\label{radial1}
    r_\text{1}=2m\left[1+\cos\left(\frac{2}{3}\arccos{\theta}\left [ -\frac{|a|}{m} \right ] \right) \right],
\end{equation}
\begin{equation}\label{radial2}
    r_\text{2}=2m\left[1+\cos\left(\frac{2}{3}\arccos{\theta}\left [ \frac{|a|}{m} \right ] \right) \right].
\end{equation}
It turns out that $r_{1}$ is associated with a rotating (prograde) orbit and $r_{2}$ with a counter-rotating (retrograde) orbit \footnote{Any particle must rotate in the same direction as the Black hole before entering the ergoregion, even if it was initially rotating in the opposite direction.}, see Figure \ref{fig:A}. 

\subsection{Shadow of the Kerr black hole}
\begin{figure}[htbp]
\centering
\includegraphics[width=0.65\textwidth]{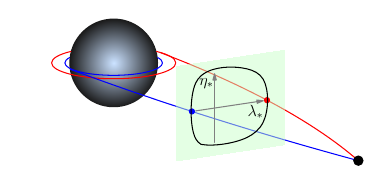}
\caption{Within the equatorial plane of a Kerr black hole, two unstable circular orbits are identified (as illustrated in the figure based on \cite{Perlick_2022}). The orbit with the smaller radius is in prograde rotation, depicted as the blue orbit, while the orbit with the larger radius is in retrograde rotation, illustrated as the red orbit. Light rays that asymptotically approach these orbits delineate two distinct points on the shadow boundary curve, as observed from the equatorial plane.}
\label{fig:A}
\end{figure}
 
\begin{figure}[htbp]
\centering
\includegraphics[width=0.7\textwidth]{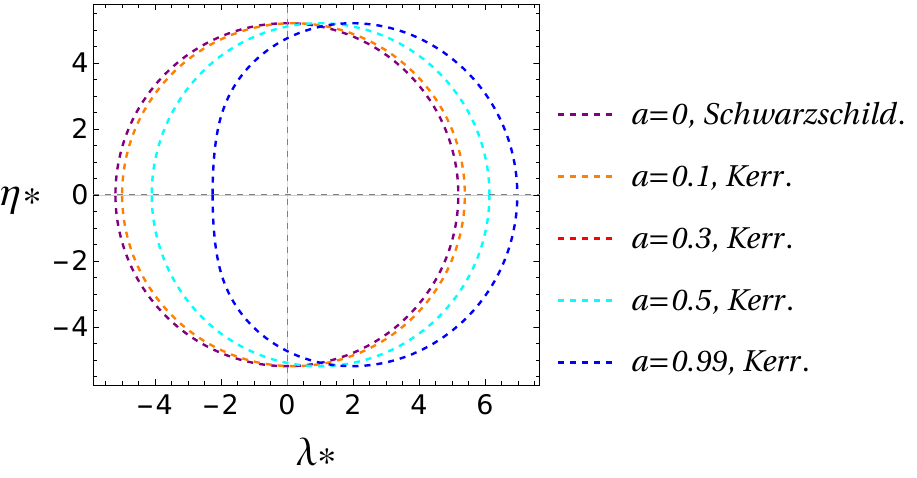}
\caption{Kerr shadow as seen by a distant equatorial observer ($r\gg m$, $\theta_{O}=\pi/2$) for different values of spin $a =0, 0.1, 0.3, 0.5, 0.99$. $\lambda*$ and $\eta*$ are the celestial coordinates. }
 \label{fig:A00}
 \end{figure}

The formation of the black hole shadow is primarily influenced by the gravitational bending of light. To analyze this phenomenon, one must consider the perspective of an observer positioned at a certain distance from the black hole~\cite{1979A&A....75..228L,10.1093/mnras/131.3.463,Bardeen:1973tla,Chandrasekhar:1985kt,Brihaye_2017,Perlick_2022,alma991019795519703414,Grenzebach_2014,Grenzebach_2015,Johannsen_2010}.
The observable shape of a black hole is defined by the boundary of its shadow. When discussing this shadow, we use celestial coordinates as outlined in Ref.~\cite{Vázquez_Esteban_2004}. Specifically, we consider a scenario where the observer's perspective aligns in such a way that the line of sight coincides with the equatorial plane of the black hole. In this configuration, the inclination angle of the observer is $\theta_{0}=\pi/2$ and the celestial coordinates are
\begin{equation}\label{celes}
    \lambda*=-\lambda,
\ \ \
    \eta*=\sqrt{\eta}.
\end{equation}
The coordinate $\lambda*$ represents the apparent horizontal distance of the image as seen from the symmetry axis, and the coordinate $\eta*$ is the apparent distance perpendicular to the equatorial plane, as illustrated in Fig.~\ref{fig:A}. Some examples of shadows for the Kerr black hole are presented in Fig.~\ref{fig:A00} for a distant equatorial observer ($r\gg m$, $\theta_{O}=\pi/2$) considering different values of spin $a =0, 0.1, 0.3, 0.5, 0.99$.

For a Schwarzschild black hole $a=0$, the shadow assumes a circular shape and achieves its largest size, characterized by an approximate radius of $3\sqrt{3}M$ in the observer's plane. With increasing spin $a>0$, the shadow size decreases, attributed to alterations in photon orbits. At elevated spin values $a>0.5$, the shadow exhibits asymmetry, manifesting itself as compression on the left side, corresponding to approaching photons, while the right side, associated with receding photons, displays elongation due to frame-dragging effects. In the particular case of Kerr black hole $a=0.99$, the shadow is markedly asymmetric and considerably smaller. The left boundary is drawn closer to the event horizon because prograde photons require less angular momentum to orbit, whereas the right boundary remains elongated since retrograde photons require greater angular momentum.

 \subsection{First-order effects of spin on the shadow of a Kerr black hole}\label{sec:srk}
In the subsequent section, we conduct an analysis of the first-order approximation in the spin parameter for the radial motion equation and subsequently we derive the expressions for $\lambda$ and $\eta$. Thereafter, a comparison is carried out to assess the influence of spin on the black hole shadow under this approximation against the equivalent findings for the exact Kerr metric. This comparative analysis substantiates the applicability of employing the first-order spin approximation in the calculations for dCS that will be elaborated later. 

The first-order rotation approximation for $R$ is expressed in the following form:
\begin{equation}
   {\textit{R}^{(1)}(r)}= E^2 r^4 - 4 a E m r \Phi - r (-2 m + r) (Q + r^2 \mu^2 + \Phi^2),
\end{equation}
where the superscript $(1)$ indicates that this quantity is valid up to first order in the rotation parameter $a$. By dividing the equation of motion corresponding to $r$ by $E^{2}$ and incorporating the reduced Carter constants $\lambda={\Phi}/{E}, \eta={Q}/{E^{2}}$, we obtain
\begin{equation}
    \mathcal{R}^{(1)}(r)=r \left[r^3 - r (\eta + \lambda^2) + 
   2 m (\eta + \lambda (-2 a + \lambda))\right].
\end{equation}
The criteria for achieving spherical orbits are delineated in Eq.~(\ref{eq:r0}), with $\mathcal{R}$ derived from Eq.~\eqref{eq:RK}.
The values of the impact parameters $\eta$ and $\lambda$ that are compatible with these conditions determine the contour of the black hole shadow.
Then, the expressions for $\eta$ and $\lambda$ take the form
\begin{align}
    \lambda^{(1)} &= \frac{(3m - r) r^2}{2 a m}, \\
    \eta^{(1)} &= \frac{12 a^2 m^2 r^2 - r^4 (3m - r)^2}{4 a^2 m^2}.
\end{align}
To satisfy the condition $\eta \geq 0$, it is essential to confine the radial coordinate within the interval specified by $r_1 \leq r \leq r_2$, where $r_1$ and $r_2$ are determined by the roots of Eq. \eqref{eqs:eta},
\begin{align}
    r_{1}^{(1)} &= \frac{1}{2} \left( 3m + \sqrt{-8 \sqrt{3} a m + 9 m^2} \right), \\
    r_{2}^{(1)} &= \frac{1}{2} \left( 3m + \sqrt{m} \sqrt{8 \sqrt{3} a + 9 m} \right).
\end{align}
\begin{figure}[htbp]
\centering

\includegraphics[width=0.305\textwidth]{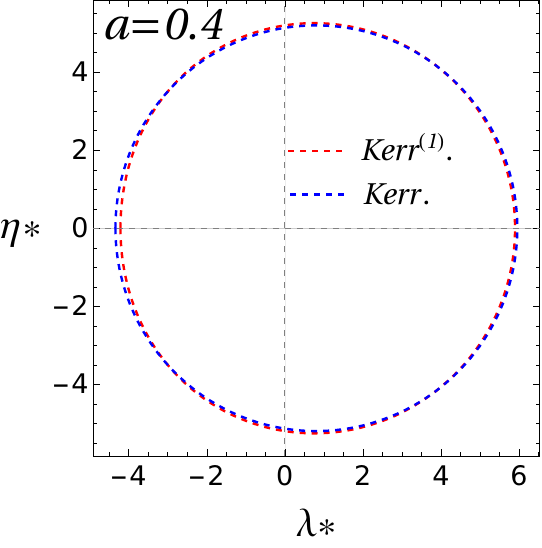} \ \ \ 
\includegraphics[width=0.3\textwidth]{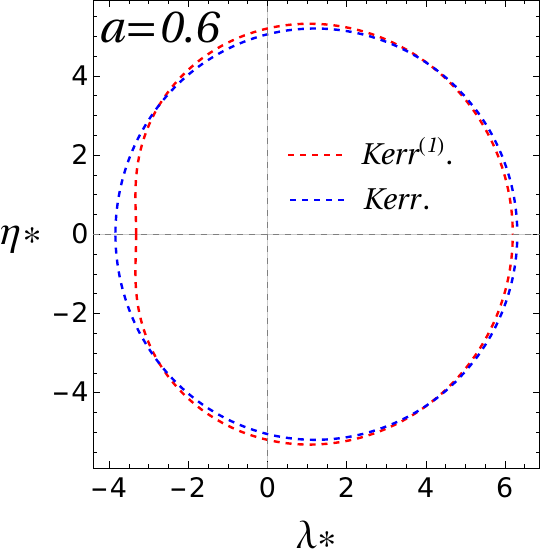}

\caption{The Kerr shadows as seen by a distant equatorial observer $\theta_{O} =\pi/2$ for different spin parameters $a = 0.4,0.6$.  
The curves are plotted with the help of equations $\lambda^*$ and $\eta^*$.  
The red dashed line represents the approximated $Kerr^{(1)}$, while the blue dashed line corresponds to the usual solution for a Kerr black hole. }
 \label{fig:FirstO}
 \end{figure}
Figure~\ref{fig:FirstO} illustrates the comparison between the black hole shadow shapes computed with the first order approximation of the Kerr metric, labeled $Kerr^{(1)}$, and the exact Kerr metric. These shadows are for $a = 0.4$ and $a= 0.6$, and are depicted for an observer positioned in the equatorial plane at a considerable distance ($r \gg m$, $\theta_{O} = \pi/2$). As the spin increases, the shadow ceases to remain centered and undergoes a horizontal shift attributable to the asymmetrical photon dynamics around the black hole. Analyzing the first-order approximation in relation to the exact Kerr solution, it is evident that for lower spin values ($a \leq 0.4$), the discrepancies between them are negligible. Conversely, at elevated spin values ($a \gtrsim 0.6$), the shadow for the Kerr linear approximation becomes markedly smaller and more displaced than predictions based on the exact Kerr metric. This observations indicate that, although the influence of spin is crucial in the study of black hole shadows, for mild values of the spin parameter the first order approximation is justified.

 \section{Black holes in dCS modified gravity: slow rotation approximation}\label{sec:hjcs}
 Adding a CS term to the Einstein-Hilbert action modifies the rotating solution of General Relativity~\cite{Jackiw_2003,Yunes_2009,Grumiller_20008,Grumiller_2008,Alexander2009,Konno_2007,Konno_2009}. The metric corresponding to the solution of the modified theory in the slow rotation approximation takes the form
  \begin{equation}\label{a5}
 ds^{2}= ds^{2}_{Kerr}+ \frac{5}{4} \frac{\alpha^{2}}{\beta \kappa} \frac{a}{r^{4}} \left(1+\frac{12}{7}\frac{m}{r}+ \frac{27}{10}\frac{m^{2}}{r^{2}} \right)\sin^{2}{\theta} d\phi dt,
 \end{equation}
 with the scalar field is given by
 \begin{equation}
     \vartheta = \frac{5\alpha a \cos{\theta}}{8\beta m r^{2}}\left( 1+\frac{2m}{r}+ \frac{18m^{2}}{5r^{2}} \right).
 \end{equation}
 In eq.~\eqref{a5}, $ds^{2}_{Kerr}$ is the slow rotation limit of the Kerr metric, $m$ is the geometrized mass of the black hole and $a$ is its specific angular momentum. 
 The coupling constants are $\alpha$ and $\beta$. This solution is valid up to the second order in the slow rotation expansion parameter ${a}/{m}$. 
 Note that this solution, derived under the dCS frame, is asymptotically flat at infinity.  
 It can be considered as a small deformation of a Kerr black hole with an additional CS scalar hair of finite energy. 
This deformation of the space-time metric is parameterized by $\alpha$
and $\beta$. When it comes to testing the theory, it is advantageous to introduce the coupling parameter
\begin{equation}
    \xi := \frac{\alpha^{2}}{\beta \kappa}.
\end{equation}
With the conventions defined in
Sec.~\ref{sec:cs}, 
 $\xi$ has units of $[\text{length}]^{4}$.   It is also customary to define
the dimensionless coupling parameter $\zeta$,
\begin{equation}
    \zeta:= \frac{\xi}{m^{4}},
\end{equation}
where $m$ is associated with the mass of the system. The rotating black hole solution to dCS gravity reduces to the Kerr metric when $\alpha=0$, or equivalently $\xi=0$ or $\zeta=0$.
To date, the best constraint on the coupling constant of the theory is given by~\cite{Yunes_2009,Ali_Ha_moud_2011,Yagi_2012}
 \begin{equation}
     \xi^{\frac{1}{4}}\leq 10^{8}\ \rm{Km}.
 \end{equation}
The corresponding dimensionless
coupling parameter is restricted to $\zeta\leq 10^{7}$ for a $10^{6}m_\odot $ black hole, and to $\zeta\leq 10^{27}$ for a $10m_\odot $ black hole.

\subsection{Hamilton-Jacobi equations in dCS and Separation of Variables.}
In the subsequent analysis, we approximate the expression in Eq.~(\ref{a5}) by neglecting the second and third terms enclosed in parentheses. This simplification is justified by observing that the influence of these terms decreases rapidly with increasing powers of $r$, specifically $r^{5}$ and $r^{6}$, rendering their contributions negligible in comparison. Therefore, we consider the metric
 \begin{equation}
 ds^{2}= ds^{2}_\text{Kerr}+ \frac{5}{4} \frac{\alpha^{2}}{\beta \kappa} \frac{a}{r^{4}} \sin^{2}{\theta} d\phi dt,
 \end{equation}
 and compute the correction to the black hole shadow to leading order. 
To analyze the general orbits of photons around the black hole, we begin by studying the separability of the Hamilton-Jacobi equation. As reviewed in previous sections, separability is possible in the case of the Kerr spacetime, using a third conserved quantity~\footnote{The other conserved quantities include the energy and the axial (z) component of angular momentum relative to infinity.}, often called Carter's constant.
For the dCS solution, we follow the same procedure. Proposing the same separable form of the Hamilton's principal function $S$ as in Eq.~\eqref{eq:hja} and focusing on null geodesics $\mu=0$, leads to
    \begin{equation}\label{a6}
        g^{tt}E^2 + g^{\phi\phi}\Phi^2 -2g^{t\phi} E \Phi +g^{rr}P_{r}^2++g^{\theta \theta}P_{\theta}^2=0.
    \end{equation}
   Equation~\eqref{a6} is presented in a general form; however, in the following it is manipulated, keeping only terms up to orders $a^2$ and $a \alpha^2$. In this way, the separation
    \begin{equation}
    f_{r}(r)=f_{\theta}(\theta)=Constant=-K,
    \end{equation}
    is achieved as in~(\ref{K}), but now 
     \begin{align}\label{a7}
    f_{r}(r) & =\frac{-P^{2}+\Delta^{2}p^{2}_{r}+\Delta r^{2}\mu^{2}}{\Delta}-\frac{5a \Delta P^{2}\alpha^{2}\Phi}{8Ekmr^{7}\beta - 4Ekr^{8}\beta}, \\
        f_{\theta}(\theta) & =-a^{2}\mu^{2}\cos^{2}{\theta}-\left ( aE\sin{\theta}-\frac{\Phi}{\sin{\theta}} \right )^{2}-p^{2}_{\theta}.
    \end{align}
    The constant $K$ can be expressed in terms of Carter's constant $Q$, which remains the same as in the Kerr case~\cite{Sopuerta_2009},
\begin{equation}
    Q= K-(aE-\Phi)^{2}.
\end{equation}
From $f_r(r) = -K$,  we get the following relation.
 
 \begin{equation}
    \Delta^{2}p^{2}_{r}=P^{2}-\Delta[Q+(aE-\Phi)^{2}+\mu^{2}r^{2}]+\frac{5a\Delta P^{2}\alpha^{2}\Phi}{8Ekmr^{7}\beta - 4Ekr^{8}\beta}.
\end{equation}

Using the relations $\dot{r}=p^{r}=g^{rr}p_{r}={\Delta} p_{r}/{\rho^{2}}$ we obtain the equation of motion for $r$ as
\begin{equation}
    \rho^{2}\dot{r}=\pm\sqrt{R(r)},
\end{equation}
where
\begin{equation}\label{a8}
    R(r)= R_{\text{K}}+\xi R_{\text{CS}}
\end{equation}
is split into the Kerr part {and} a dCS correction,
\begin{align}
    R_{\text{K}}&= P^{2}-\Delta[Q+(aE-\Phi)^{2}+\mu^{2}r^{2}], \\
    R_{\text{CS}}&=\frac{5a\Delta P^{2}\Phi}{8Emr^{7} - 4Er^{8}}. 
\end{align}
On the other hand, equation $f_\theta(\theta) = -K$ leads to
\begin{equation}
    K=a^{2}\mu^{2}\cos^{2}{\theta}+\left ( aE\sin{\theta}-\frac{\Phi}{\sin{\theta}} \right )^{2}+p^{2}_{\theta},
\end{equation}
From this relation it follows that we must have $K\geq 0$. The momentum in terms of Carter's constant becomes 
\begin{equation}
    p^{2}_{\theta}=Q-\cos^{2}{\theta}\left ( a^{2}(\mu^{2}-E^{2})+\frac{\Phi^{2}}{\sin^{2}{\theta}} \right ).
\end{equation}
Using $\dot{\theta}=p^{\theta}=g^{\theta\theta}p_{\theta}={p_{\theta}/\rho^{2}}$ 
we can obtain the equation of motion for $\theta$,
\begin{equation}
    \rho^{2}\dot{\theta} = \sqrt{Q-\cos^{2}{\theta}\left ( a^{2}(\mu^{2}-E^{2})+\frac{\Phi^{2}}{\sin^{2}{\theta}} \right ) } = \sqrt{\Theta(\theta) }.
\end{equation}
We must notice that by replacing $\xi = 0$ in Eq.~\eqref{a8} we recover the expression for the Kerr metric (to the second order in $a$). Also, it is worth noticing that the function $\Theta(\theta)$ is the same as for the Kerr geometry.%

\subsection{Shadows of dCS black holes}

The conditions for the spherical orbits are given by Eq.~(\ref{eq:r0}), with $\mathcal{R}$ constructed out of Eq.~\eqref{a8}.

\begin{figure}[htbp]
\centering

\includegraphics[width=0.3\textwidth]{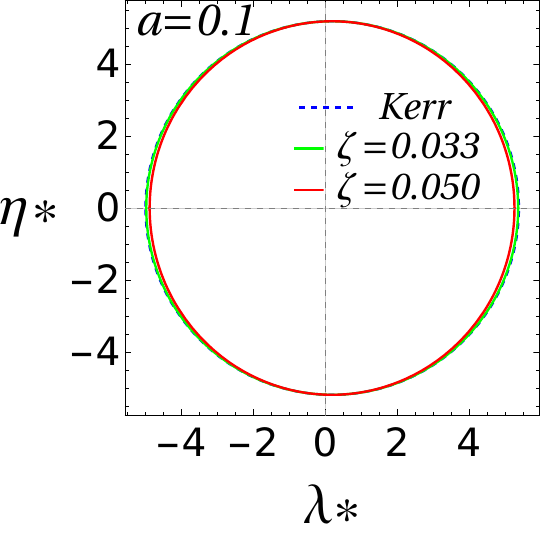}
\includegraphics[width=0.3\textwidth]{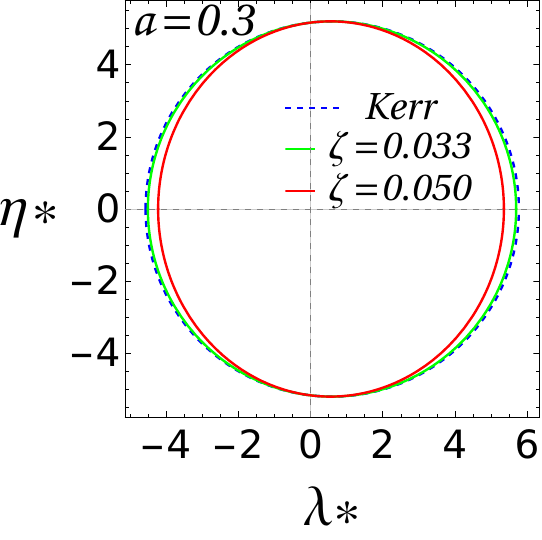}
\includegraphics[width=0.3\textwidth]{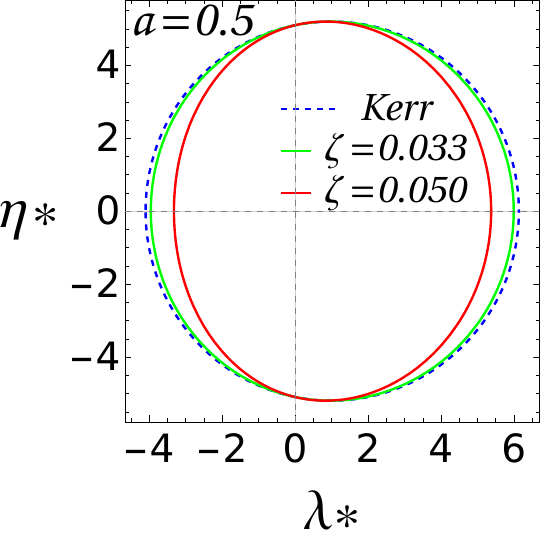}
\caption{ The dCS shadows observed by a distant equatorial observer $\theta_{O} =\pi/2$ have been analyzed for varying spin parameters $a = 0.1,0.3,0.5$ and distinct values of $\zeta = 0.033, 0.050$. The corresponding curves are derived using equations $\lambda*$ and $\eta*$. }
 \label{fig:A3}
 \end{figure}

The values of the impact parameters $\eta$ and $\lambda$ that are compatible with these conditions determine the contour of the black hole shadow. In the case of slow rotating dCS black holes, the parameters $\eta$ and $\lambda$ compatible with the Eq. \eqref{eq:r0} belong to two possible families, as in the case of Kerr geometry. However, one of these families is not consistent with the conditions that must be satisfied by the function $\Theta(\theta)$~\cite{Chandrasekhar:1985kt}. In fact, the family of allowed parameters is the one that in the limit $\alpha= 0$ coincides with the solution for the Kerr geometry. Then, the expressions for $\eta$ and $\lambda$ take the form
\begin{align}
    \label{lamf}
    \lambda & =\lambda_\text{K}+\xi\lambda_\text{CS}, \\
    \label{etaf}\eta & =\eta_\text{K}+\xi\eta_\text{CS},
\end{align}
where 
\begin{align}
    \lambda_\text{CS} & = \frac{ 5\left( a^{2}+rf_{2}\right )^{2} \left (r^{2}f_{3}+a^{2}f_{-1}\right) \left(3rf_{2}^{2} +a^{2}\left ( 3r+f_{7} \right ) \right)}{2ar^{7} f_{-2}^{2}f_{1}^{3}},\\
    \eta_\text{CS} & = \frac{5\left (a^{2}-r f_{2} \right)^{2}\left ( r^{2}f_{3}+a^{2}f_{-1}\right ) \left ( 
3rf_{3}f_{2}^{2}+a^{2}f_{23}^{*} \right )}{a^{2} r^{5}\left ( -f_{1} \right )^{4}f_{2}^{2}} ,
\end{align}
and the functions $f$ are defined as follows:
\begin{align}
     f_{n} & =r-nm, \\ 
      f_{n}^{*} & =\left (1-n \right )rm+nm^{2}+5r^{2}, \\
       f_{n}^{**} & =\left ( 1-n \right )rm+\left ( 
n-1 \right )m^{2}+r^{2}.
\end{align}
Here, we also define $f_n^{**}$ for future use. 
To ensure that condition $\eta \geq 0$ is satisfied, we need to constrain the radial coordinate to the range of $R_1 \leq r \leq R_2$, where $R_1$ and $R_2$ are determined by the roots of Eq. \eqref{etaf}, as 
\begin{align}
    R_1&=r_\text{1}+\xi r_\text{CS1}, \label{eq:cs1}\\
    R_2&=r_\text{2}+\xi r_\text{CS2}\label{eq:cs2},
\end{align}
with 
\begin{align}
 r_\text{CS1} &=\frac{ 25\left(a^{2}+r_\text{1}f_{2} \right)^{2}\left ( r_\text{1}^{2}f_{3}+a^{2}f_{-1}\right )\left( a^{2}f_{23}^{*}-r_\text{1}f_{3}f_{2}^{2} \right)}{4r^{7}_\text{1}f_{1}f_{2}^{2}f_{3} \left ( r_\text{1}f_{4}^{**}-a^{2}m \right )}, \\
 r_\text{CS2} &=\frac{ 25\left(a^{2}+r_\text{2}f_{2} \right)^{2}\left ( r_\text{2}^{2}f_{3}+a^{2}f_{-1}\right )\left( a^{2}f_{23}^{*}-r_\text{2}f_{3}f_{2}^{2} \right)}{4r^{7}_\text{2}f_{1}f_{2}^{2}f_{3} \left ( r_\text{2}f_{4}^{**}-a^{2}m \right )}.
\end{align}
In Eqs.~\eqref{eq:cs2} and \eqref{eq:cs2}, $r_{1,2}$ are the GR limits for the radial coordinate given in Eq.~\eqref{radial1} and $\eqref{radial2}$.
The allowed values of the parameters $\eta$ and $\lambda$ determine the shadow of the black hole in dCS modified gravity. If a black hole is located between a light source and an observer, some photons reach the observer after being deflected by the gravitational field of the black hole. However; those with small impact parameters end up falling into the black hole without reaching the observer.
In Fig.~\ref{fig:A3} we show a representation of dCS shadows based on equations~\eqref{celes}, \eqref{lamf} and ~\eqref{etaf}, for the equatorial observer angle $\theta_{O} = {\pi}/{2}$, spin parameters $a = 0.1, 0.3, 0.5$, and the values of the parameter $\zeta = 0.033, 0.050$. The generated curves illustrate the influence of these parameters on the observed shadows. 

 \subsection{Images of dCS black hole with accretion disk} 
 
\begin{figure}[htbp]
\centering
\includegraphics[width=0.3\textwidth]{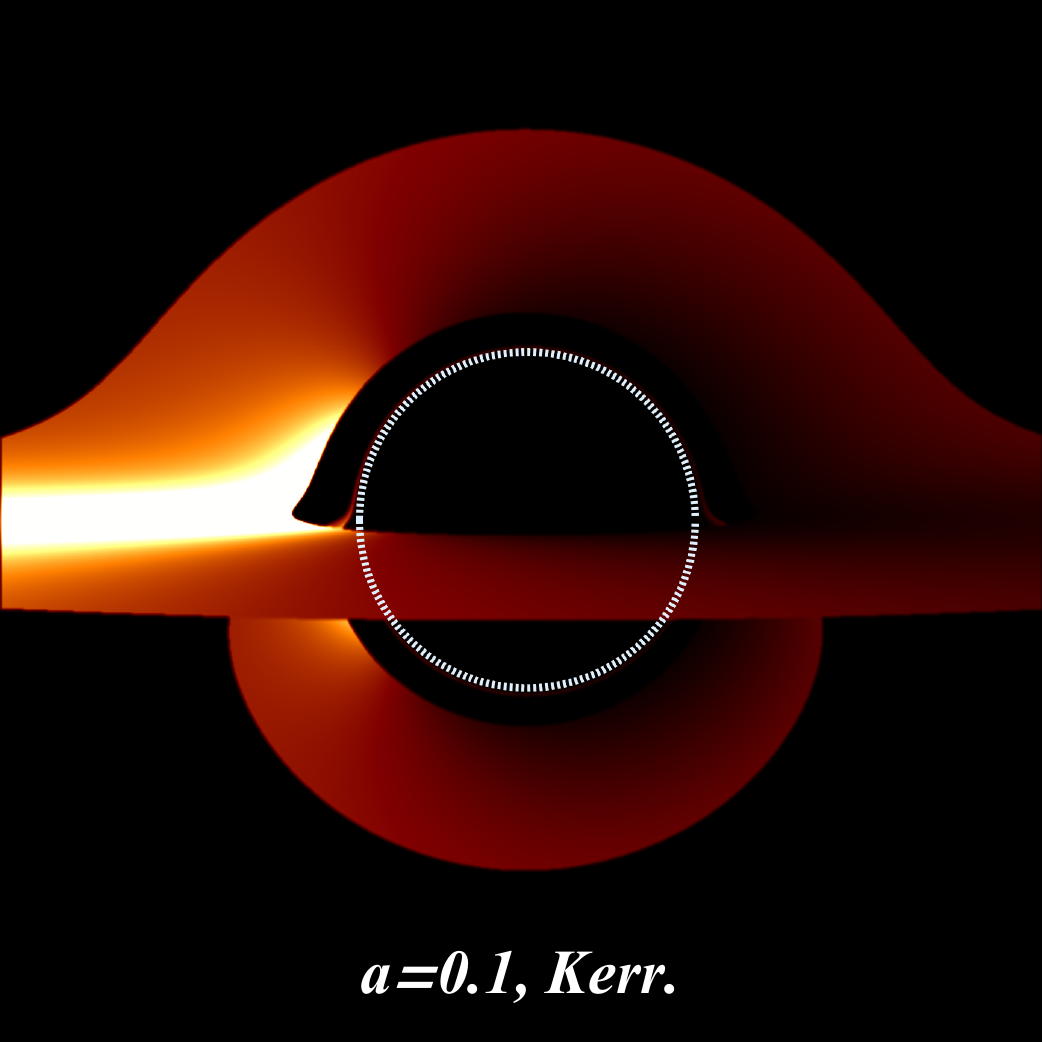}
\includegraphics[width=0.3\textwidth]{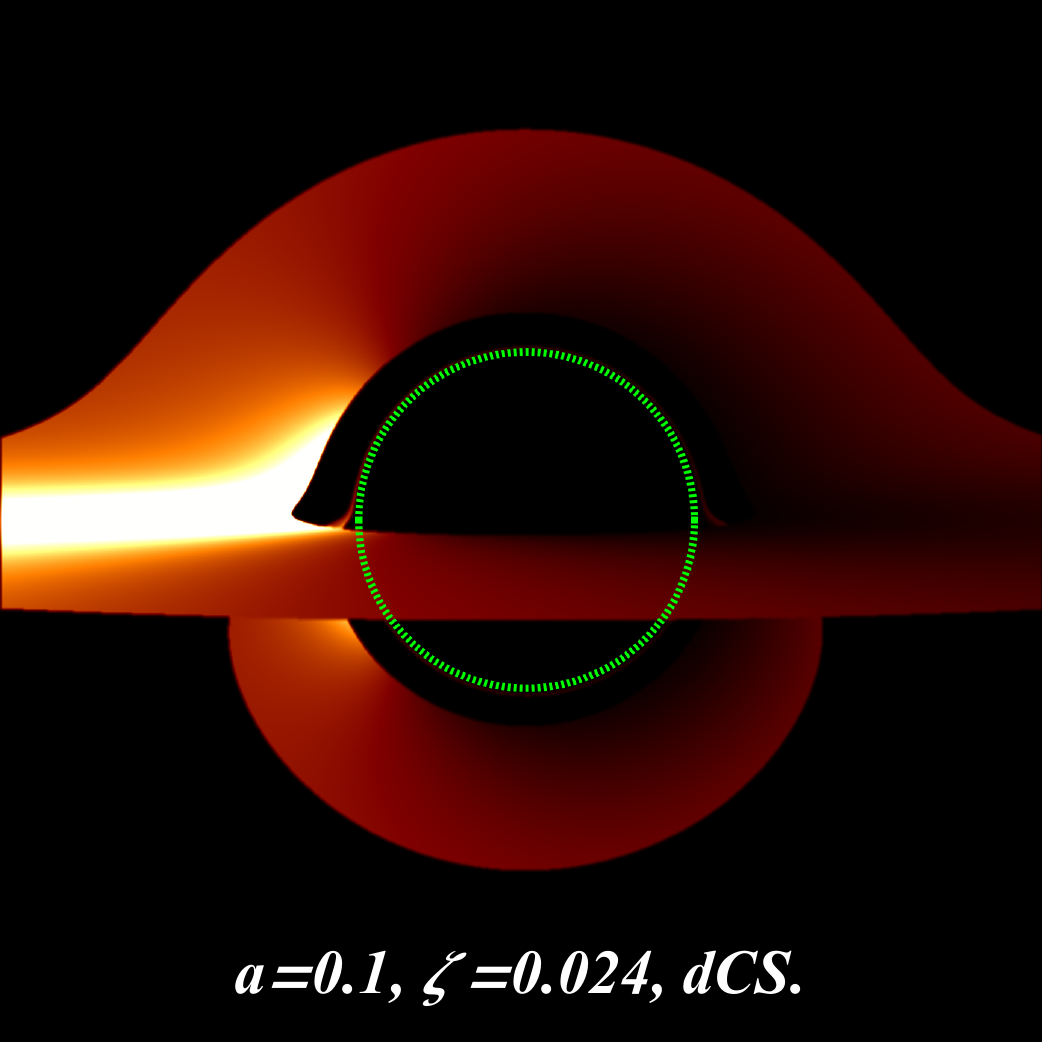}  
\includegraphics[width=0.3\textwidth]{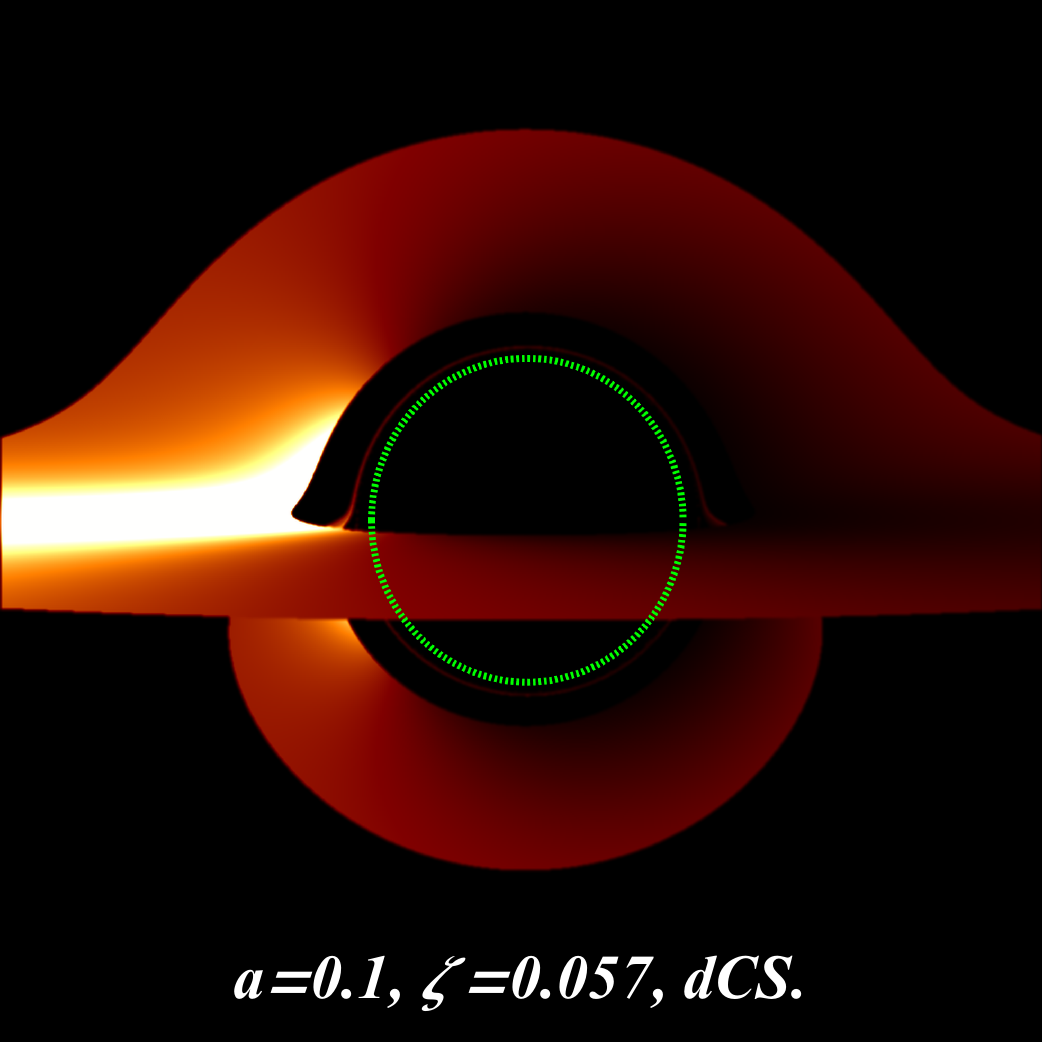}
\caption{The image presents a sequence of black hole shadow visualizations, progressing from left to right. The first panel depicts the shadow of a Kerr black hole with spin parameter $a = 0.1$, outlined by a white dotted line, superimposed on the visualization of a geometrically thin accretion disk in the equatorial plane ($\theta = \pi/2$). The second and third panels illustrate the black hole shadow in the dynamical Chern-Simons (dCS) gravity framework, also for $a = 0.1$, outlined by a green dotted line and overlaid on the corresponding accretion disk images. These visualizations are generated using the Gyoto ray-tracing code~\cite{Vincent_2011}.}
\label{fig:Cskdisc}
\end{figure}

 Accretion disks, essential elements situated in the equatorial plane of black holes, play a crucial role in the emission of observable light. These dynamic structures are characterized by a continuous flow of matter. The behavior of matter within the accretion disk is significantly influenced by its proximity to the black hole. To create these images, we use a thin accretion disk, as described in Ref.~\cite{1974ApJ...191..499P}. This model simplifies the analysis of accretion processes by assuming that the disk is geometrically thin compared to its radial extent. In addition, the disk is considered stationary and axisymmetric. These simplifications allow us to describe a two-dimensional disk using few important parameters such as the mass and spin of the black hole, the dimensionless coupling parameter of dCS, $\zeta$, the radius of the ISCO that determines the smallest stable orbits of massive particles, the geometrical thickness of the disk in geometric units, the angle (in radians) between spin axis and disk surface, and the position of observer angle that we fix to $\theta_{O}= \pi/2$.
In Figure~\ref{fig:Cskdisc}, we present thin accretion disks surrounding both a Kerr ($a=0.1$) and a dCS ($a=0.1$, $\zeta = 0.024, 0.057$) generated numerically, with superimposed images of null geodesics (dotted lines) of a photon sphere, obtained using the Hamilton-Jacobi formalism. At the given resolution of the figure (1000 × 1000 pixels), the images of Kerr and dCS for $\zeta = 0.024$ black holes are indistinguishable from the naked eye. Any disparities between the two images are on the order of pixel size. On the other hand, for $\zeta = 0.057$, differences emerge between the analytically generated geodesic and the numerically generated background. Considering our approximation scheme, it is important to note that the precision of the analytic solutions decreases for higher values of $\zeta$.

These images were generated using the \verb 'Gyoto' code, an open-source ray tracing tool~\cite{Vincent_2011}. \verb 'Gyoto' operates by numerically integrating null geodesic equations in a backward fashion, commencing from the observer's screen. This technique allows for the computation of visual representations of the accretion disk around the black hole \footnote{Let us remember that the angular size of the black hole shadow depends only on gravity effects, not on the astrophysical assumptions we made on the disk structure and emission.}.

 \subsection{Observables for the shadow}
 
\begin{figure}[htbp]
\centering
\includegraphics[width=0.6\textwidth]{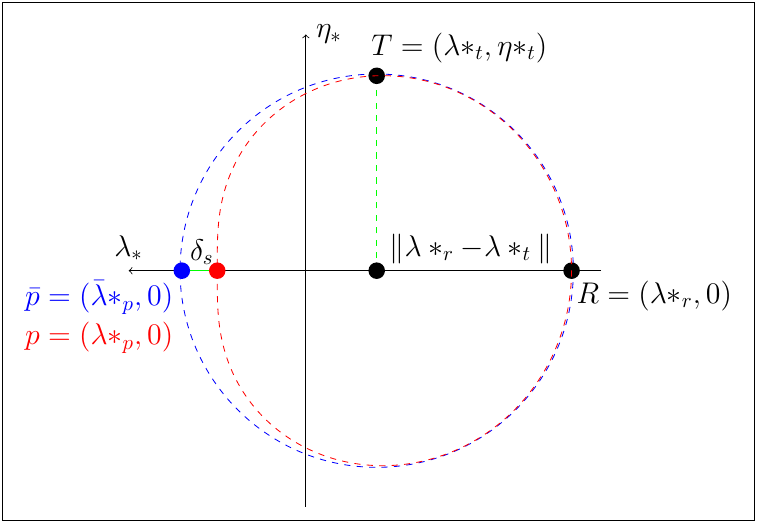}
\caption{The observable $\delta s$ quantifies the extent to which the actual shadow deviates from the reference circle,  where $(\bar{\lambda}*_{p},0)$ and $(\lambda*_{p},0)$ represent points where the reference circle and the shadow contour intersect the horizontal axis, positioned on the opposite side of $(\lambda*_{r},0)$.}
\label{obser}
\end{figure}

In this section, we analyze quantitatively the difference
between the shadow of rotating black holes in dCS theory and in GR.  
To this end, we compute the shadow radius $R_\text{s}$ and the distortion parameter $\delta_\text{s}$. These observables enable us to perform a precise exploration of the deviations with respect to the Kerr shadow. Our methodology is based on the well-established framework provided in Refs.~\cite{Hioki_2009,Wei_2019,Lee_2021}. 
The shadow radius $R_\text{s}$,
defined in Ref.~\cite{Hioki_2009} as 
\begin{equation}
 R_\text{s}=\frac{\left ( \lambda*_{t}-\lambda*_{r} \right )^{2}+\eta*^{2}_{t}}{2 \left | \lambda*_{r}-\lambda*_{t} \right|},
\end{equation}
characterizes the radius of a reference circle. The quantities in the definition above correspond to the points $T$ and $R$ shown in Fig.~\ref{obser}, whose coordinates are
$(\lambda*_{t},\eta*_{t})$ and $(\lambda*_{r},0)$, respectively.

\begin{figure}[htbp]
\centering
\includegraphics[width=0.48\textwidth]{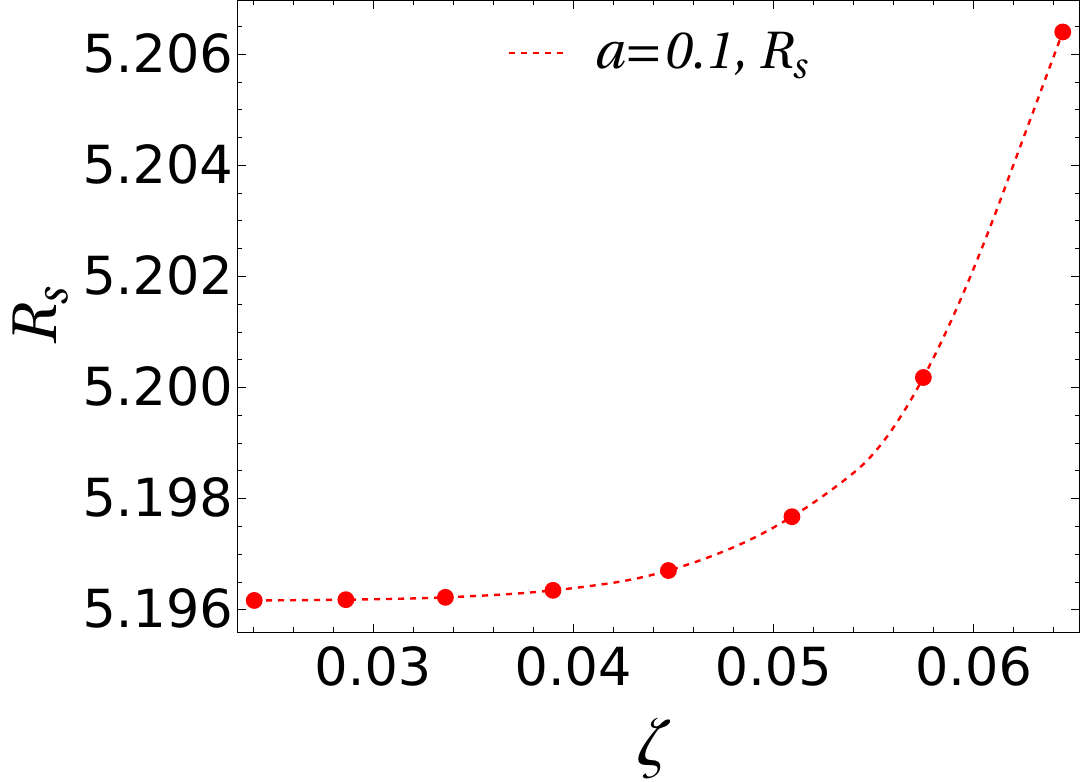}
\includegraphics[width=0.48\textwidth]{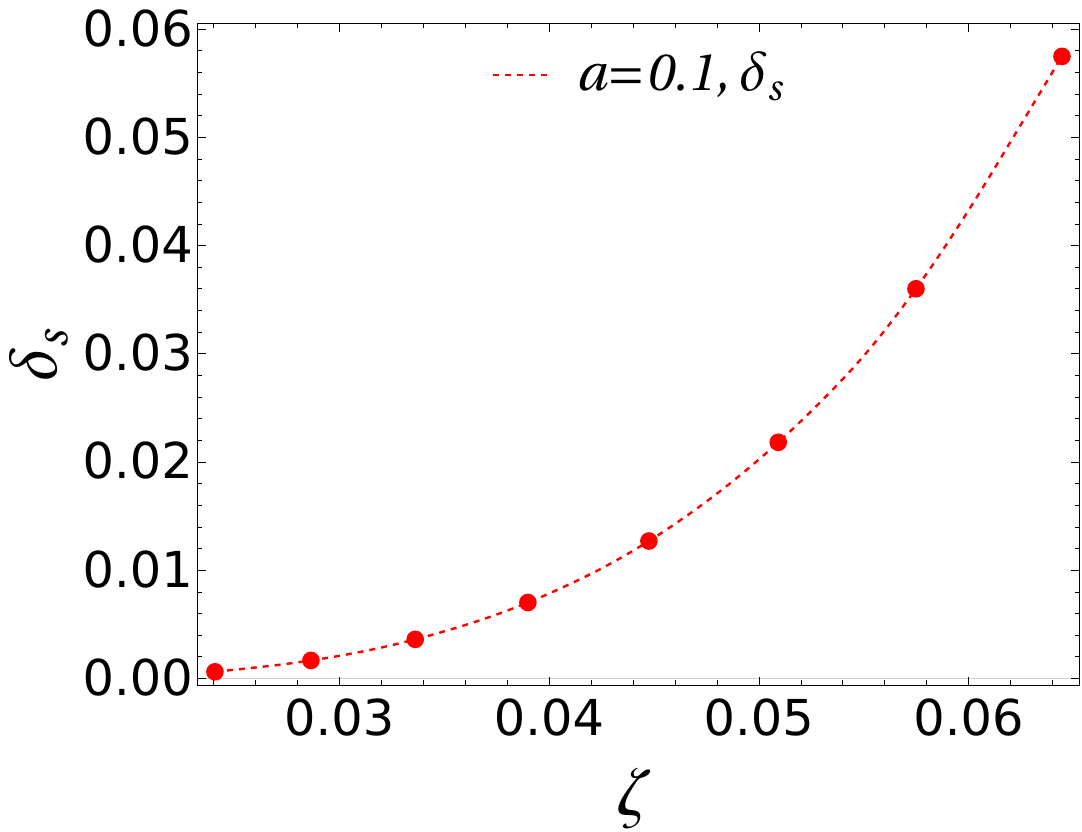}
\caption{The observables $R_{s}$ and $\delta_s$ as functions of the dCS coupling parameter $\zeta$, corresponding to the shadow of a black hole positioned at the origin of coordinates and in the equatorial plane, with spin parameter $a = 0.1$. 
As $\zeta$ increases, the shadow becomes more distorted.}
\label{obs1}
\end{figure}

On the other hand, the distortion parameter $\delta_{s}$ quantifies the extent to which the actual shadow deviates from the reference circle and is defined by
\begin{equation}   \delta_\text{s}=\frac{ \left | \bar{\lambda}*_{p}-\lambda*_{p} \right |}{R_\text{s}},
  \end{equation}
 where $(\bar{\lambda}*_{p},0)$ and $(\lambda*_{p},0)$ represent points where the reference circle and the shadow contour intersect the horizontal axis, positioned on the opposite side of $(\lambda*_{r},0)$, as shown in Fig.~\ref{obser}. 
The observables $R_s$ and $\delta _s$ as functions of the dimensionless coupling parameter dCS $\zeta$ are shown in Fig.~\ref{obs1}. 

\begin{figure}[htbp]
\centering
\includegraphics[width=0.4\textwidth]{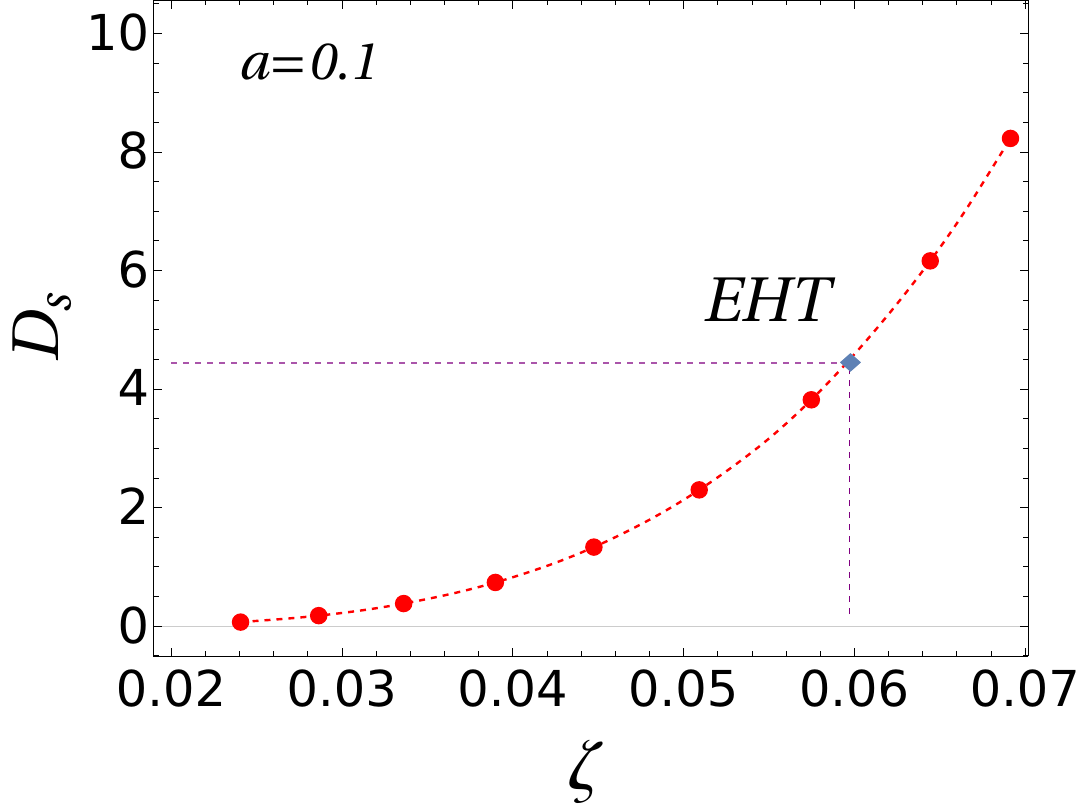}
\includegraphics[width=0.4\textwidth]{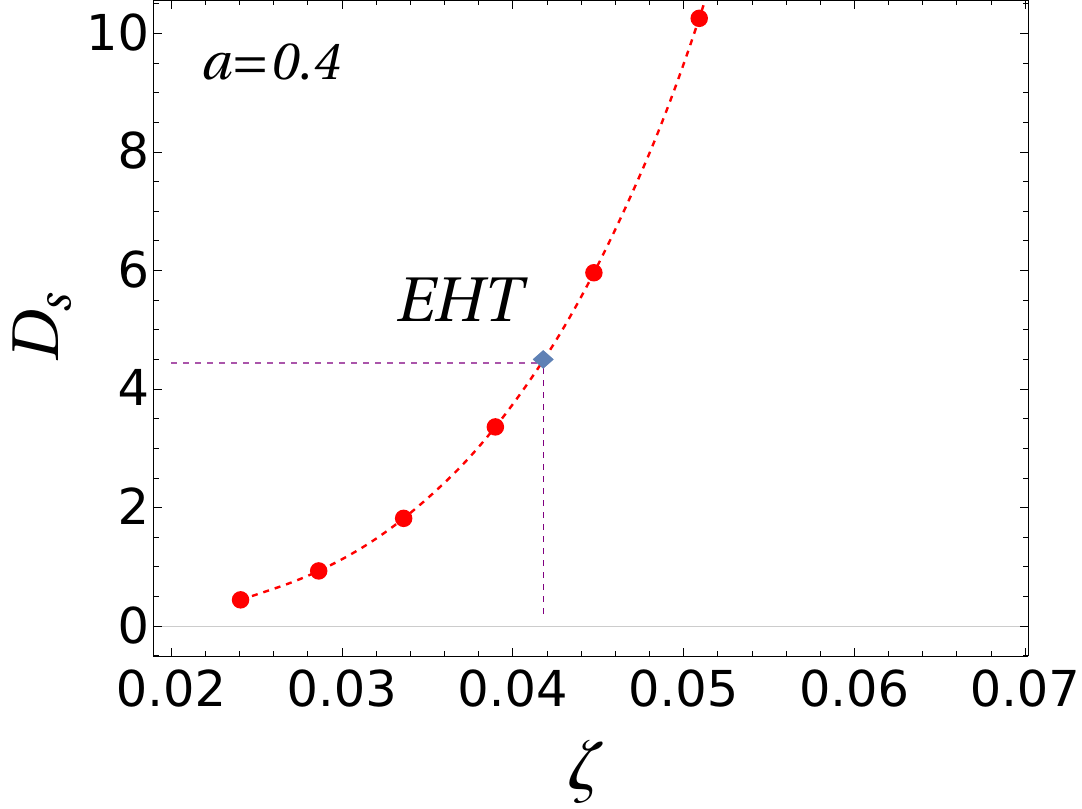}

\caption{The relative difference $D_{s}$ with respect to a Kerr black hole measured from $(\lambda*_{r},0)$ as functions of the dCS coupling parameter $\zeta$, corresponding to the shadow of a black hole positioned at the origin of coordinates in the equatorial plane $\vartheta_{O} = \pi/2$ and the spin parameter $a = 0.1$ (left) and $a=0.4$ (right). The validity of the slow rotation approximation for these values of $a$ is justified by the discussion in Sec.~\ref{sec:srk}.}
\label{obs2}
\end{figure}

The applicability of these observables is constrained to particular shadow configurations~\cite{Abdujabbarov_2015}, necessitating certain symmetrical properties and potentially proving inadequate for black holes governed by specific modified gravity theories. For dCS, from the results of the previous section, we see that the coordinate $\lambda*_r$ is significantly affected by the value of the CS coupling parameter. Therefore, we propose an additional parameter to quantify this deformation, designated as the ``relative difference'' $D_\text{s}$. This parameters compares the position of $\lambda*_{rCS}$ to that of a reference Kerr black hole, $\lambda*_{r}$, \textit{i.e.},
\begin{equation}
    D_\text{s}=100\left | \frac{\lambda*_{r}-\lambda*_{rCS} }{ \frac{\lambda*_{r}+\lambda*_{rCS}}{2}}\right |,
    \label{eq:Ds}
\end{equation}
where both the CS and the Kerr solutions are considered with the same spin parameter. The estimation of this observable is reported in Figure~\ref{obs2} for spin parameters $a = 0.1$ and $a = 0.4$. In the same figure, we include the relative difference allowed by the uncertainty in the shadow radius of $SgrA^{*}$, represented by a square marker. This reference point is obtained from Eq.~\eqref{eq:Ds} by assigning $\lambda*_{r}$ to the central value of the $SgrA^{*}$ shadow radius and $\lambda*_{rCS}$ to the maximum value permitted by the observational uncertainty. This estimation provides an indication of how tightly the parameter $\zeta$ can be constrained given the current precision of EHT measurements.

It is important to note that we are only incorporating the uncertainty range reported by EHT; we are not performing a direct comparison with the actual observations of $SgrA^{*}$, which would require knowledge of the dCS solution beyond the slow-rotation regime, particularly for  $a \gtrsim0.7$~\cite{Daly:2023axh,Dokuchaev:2023obv}, where the approximation used here ceases to be valid. Nevertheless, Table~\ref{tab:zeta_values} reveals that as the spin parameter increases, the maximum allowed value of $\zeta$ decreases. Consequently, large values of $\zeta$ are excluded, as they would induce distortions incompatible with current EHT observations.\footnote{A similar estimation can be carried out using the shadow of $M87^{}$; however, it results in a weaker constraint on $D_\text{s}$.}
This shows that the relative difference $D_\text{s}$ serves as a means to quantify the deviations within the shadow. These deviations are directly attributed to variations in the coupling constants $\alpha$ and $\beta$ associated with the scalar field in the context of dCS modified gravity. 
\begin{table}[htbp]
    \centering
    \begin{tabular}{c c c c c }
        \hline
        $a$ & 0.1 & 0.2 & 0.3 & 0.4 \\
        \hline
        $\zeta_{max}$ & 0.0597 & 0.0499 & 0.0446 & 0.0418 \\
        \hline
    \end{tabular}
    \caption{Maximum $\zeta$ values corresponding to various spin parameters, considering the uncertainty in EHT measurements for $Sgr A^{*}$.}
    \label{tab:zeta_values}
\end{table}

This analysis demonstrates that the relative difference $D_\text{s}$ serves as an effective diagnostic tool for quantifying deviations in the shadow profile. These deviations originate from variations in the coupling constants $\alpha$ and $\beta$ associated with the scalar field in the context of dCS modified gravity.

\section{Discussion  and conclusions}\label{sec:con}
In this work, we explore the behavior of null geodesics around a slowly rotating black hole within the framework of dCS modified gravity, specifically in scenarios characterized by a small coupling constant. An important observation in our analysis is the separability of the Hamilton-Jacobi equation. However, for certain black hole solutions, the equation does not readily permit variable separation. To address this, we use an approach based on the hierarchy of expansion orders in inverse powers of the radial coordinate, which facilitates the identification of the most significant modifications to the geodesic equations and allows us to systematically neglect some terms from the equations of motion.  

From these equations, we find that the shape of a black hole shadow in dCS modified gravity depends primarily on the parameters $m$, $a$, and $\zeta$. The mass parameter determines the size of the shadow, while $a$ and $\zeta$ influence the deviations from perfect circularity. This implies that the shadow's appearance becomes a distinguishing factor between Kerr geometry and its dCS modification. In this alternative theoretical framework, for a fixed rotation parameter $a$, the shadow consistently appears larger and more distorted than predicted by GR, with the distortion increasing progressively for higher values of $\zeta$, mainly due to the substantial influence of the dCS term.

EHT findings indicate that the spin parameter for $SgrA^{*}$ remains uncertain but  lie in the range \( a \gtrsim 0.7 \)~\cite{Daly:2023axh,Dokuchaev:2023obv}, with the photon ring exhibiting a diameter of $51.8 \pm 2.3\mu{\rm as}$. To constrain the dCS coupling parameter, we examined the extent to which deviations induced by dCS gravity can alter the shadow radius while remaining consistent with the observational uncertainties reported by the EHT. Since a meaningful comparison requires both Kerr and dCS black holes to share the same spin parameter, our analysis is limited to the regime \( a \lesssim 0.4 \), where the slow-rotation approximation remains valid for the dCS solution.

Our results show that the allowed values of the dimensionless coupling parameter $\zeta$ depend on the black hole’s spin. Specifically, we find that for $a = 0.4$, the upper bound is $\zeta_{\text{max}} \sim 0.042$, while for $a = 0.1$, it increases to $\zeta_{\text{max}} \sim 0.06$. Extrapolating from this trend in Table~\ref{tab:zeta_values}, we estimate that for a rapidly rotating black hole with a $\sim 0.7$—closer to the expected spin of $SgrA^{*}$—the constraint could tighten to $\zeta_{\text{max}} \sim 0.03$.

These projections underscore the potential of black hole shadow observations as a diagnostic tool for probing deviations from General Relativity. As observational precision improves, future EHT data will enable more stringent tests of alternative gravity theories such as dCS. In this way, continued scrutiny of black hole shadows offers a promising avenue for refining our theoretical models and constraining the space of viable modifications to Einstein’s theory.

\section{Acknowledgments}
The authors appreciate the support provided by CONAHCyT through grants 932448 and DCF-320821, which made this research possible.

\bibliography{ref}

\end{document}